\documentclass[12pt,twoside,a4paper,fleqn]{article}
\usepackage[left=25mm,top=20mm,right=14mm,bottom=30mm]{geometry}

\usepackage{epsfig}
\usepackage{graphicx}
\usepackage{amssymb}
\usepackage{mathrsfs}
\usepackage{dcolumn}
\usepackage{amsmath}
\usepackage{multirow}
\usepackage{multicol}

\newcommand{\pr}{\partial}

\newcommand{\rta}{\rightarrow}

\newcommand{\upa}{\uparrow}
\newcommand{\dwa}{\downarrow}
\newcommand{\ep}{\epsilon}

\newcommand{\p}{\prime}
\newcommand{\om}{\omega}
\newcommand{\ra}{\rangle}
\newcommand{\la}{\langle}

\newcommand{\beq}{\begin{equation}}
\newcommand{\eeq}{\end{equation}}

\newcommand{\pp}{{\p\p}}

\newcommand{\ball}{\begin{align}}
\newcommand{\eall}{\end{align}}

\newcommand{\beqar}{\begin{eqnarray}}
\newcommand{\eeqar}{\end{eqnarray}}

\newcommand{\twop}{{\prime\prime}}

\newcommand{\ben}{\begin{enumerate}}
\newcommand{\een}{\end{enumerate}}
\newcommand{\mud}{\mu_{d}}
\newcommand{\mus}{\mu_{s}}
\makeatletter
\newcommand*{\rom}[1]{\expandafter\@slowromancap\romannumeral #1@}
\makeatother

\begin{document}
\title{A theory of resistivity in Kondo lattice materials: the memory function approach}
\author{ Komal Kumari$^*$, Raman Sharma$^*$ and Navinder Singh$^{**}$ \\ $^*$Department~of~Physics,~Himachal~Pradesh~University,\\~Shimla,~India, Pin:171005.\\$^{**}$Physical Research Laboratory, Ahmedabad,\\ India, Pin: 380009.
	\footnote{~Email:~sharmakomal611@gmail.com, komal.phyhpu@gmail.com}
	\footnote{~Email:~raman.sharma@hpuniv.ac.in, }
	\footnote{ ~Email:~navinder@prl.res.in}}
\date{17/12/2019}
\maketitle

\begin{abstract}
We theoretically analyse D.C. resistivity($\rho$) in the Kondo-lattice model using the powerful memory function approach. The complete temperature evolution of $\rho$ is investigated using the W\"{o}lfle-G\"{o}tze expansion of the memory function. The resistivity in this model originates due to spin-flip magnetic scattering of conduction $s$-electron off the quasi-localized $d$ or $f$ electron spins. We find the famous resistivity upturn at lower temperature regime ($k_{B}T<<\mud$), where $\mud$ is the effective chemical potential of $d$-electrons. In the high temperature regime $(\mud<<k_{B}T)$ we discover that $\rho \propto T^{\frac{3}{2}}$. The worked out theory is quantitatively compared with experimental data and reasonably good agreement is found.
\end{abstract}

\section{Introduction}

In heavy fermion materials such as $Ce Cu_{2}Si_{2}$, $ Ce Cu_{2}Ge_{2}$, $U Ru_{2}Si_{2}$, $U Pd_{2}Al_{3}$ etc. \cite{patrik,yang} and also in nano-scale granular aluminum \cite{bachar} at sufficiently high temperatures $(T>T_{K})$, where $T_{K}$ is the Kondo temperature) it becomes possible to divide the electronic system into two components: (1) mobile or conduction $s$-electrons, and (2) localized d or f moments. The localized moments leads to the typical curie susceptibility ($\chi\propto\frac{1}{T}$) and the itinerant $s$-electrons provide the electrical conduction. As temperature is lowered the conduction electron spins start to quantum mechanically hybridize with the localized $d$ or $f$-moments. At sufficiently low temperature $ T_{K}$, conduction electrons and localized $f$ moments form what is known as Kondo singlets. The process of hybridization is gradual one starting from higher temperature where $f$-moments are free, to very low temperature ($T<<T_{K}$) where $f$-moments form spin singlets with conduction $s$ electrons. In this very low temperature regime emerges "hybridized electrons". These "hybridized electrons" are very heavy (their mass is many order of magnitude larger than free electron electron mass). Thus these systems are called heavy Fermion systems. It turns out that the Fermi volume contains both the conduction $s$-electrons and the "localized" $f$ or $d$ electrons in the $T<T_{K}$, and the superconducting transition happens in these "heavy electrons"\cite{pf}. However, in the high temperature regime Fermi volume contains only the conduction electrons (not the localized $f$ or  $d$-moments)\cite{patrik,yang}.

The current investigation is devoted to a different problem of electrical conduction in such systems. In the current investigation which is valid for $T>T_{K}$  we study the scattering of conduction $s$-electrons via the quasi-localized  $f$-moments. Our aim is to calculate the temperature dependence of the electrical resistivity originating from magnetic scattering.

Resistivity from magnetic scattering is a well know phenomenon. The Kondo effect of resistivity minimum in materials containing magnetic impurities such as $AuFe$  is  well studied\cite{schloarpedia,kondo,hewson}. It occurs due to spin flip scattering of conduction electrons via spin flips of localized magnetic impurity spin. J. Kondo explained it using second order perturbation theory \cite{schloarpedia,kondo,hewson}. In other words it takes into account the spin flip of the impurity and scattering electron as an intermediate state:
\beqar
 \sum_{k^\p} J(k\dwa,\upa\rta k^{\pp}\upa,\dwa).J( k^{\pp}\upa,\dwa\rta k^\p\dwa,\upa)\frac{(1-f_{k^{\pp}})}{\ep_{k}-\ep_{k^\p}}, \label{kon1}
 \eeqar
where the factor $1-f_{k^{\pp}}$ represents the probability that the state $| k^\twop\ra$ is empty.  The above term represents the scattering of an electron with wavevector $k$ and spin state $|\dwa\ra$ and the impurity in spin state $|\upa\ra$  into an intermediate state having electron with wavevector $k^\twop$ but flipped spins for both the impurity and the electron. Then from this intermediate state electron scatters to a final state with wavevector $k^\p$ with one more flips of electron and impurity spins, such that the spin states returns back to its original form. As is well known the resistivity due to above Kondo term scales as $log(T)$ \cite{schloarpedia,kondo,hewson,shenoy}. However, this calculation does not capture full temperature evolution of resistivity. Our calculation using memory function formalism incorporates the above Kondo term and we analytically obtain the full temperature dependence of the resistivity including the high temperature behaviour($\rho\propto T^{\frac{3}{2}}$). In our calculation 
 the coupling of $s$-electrons with quasi-localized $d$ or $f$-moments is taken to be the Kondo Coupling.  We treat d or f electrons as quasi-localized instead of perfectly localized ones as considered in the standard Kondo problem. Perfect localization of f or d electrons occurs in the integer valence compounds (at half filling)\cite{patrik}. Due to integer valence and strong  onsite Coulomb repulsion (Hubbard U) double occupancy at a given site is prohibited.  In our calculation we consider systems away from integer valence and d or f electrons are treated as quasi-localized, and they form a small Fermi surface ( refer to section 2 and appendix A).  The coupling Hamiltonian is the Kondo lattice Hamiltonian also known in the literature as $s$-$d$ Hamiltonian:
\beqar
H_{sd}=\frac{J}{N}\sum_{k^{\p}k} \bigg\{ a^{\dagger}_{k^{\p}\uparrow} a_{k\downarrow}S^{-}(k^{\p}-k)+ a^{\dagger}_{k^{\p}\downarrow} a_{k\uparrow}S^{+}(k^{\p}-k)+(a^{\dagger}_{k^{\p}\uparrow}
a_{k\upa}-a^{\dagger}_{k^{\p}\downarrow} a_{k\dwa})S^{z}(k^{\p}-k)\bigg\} \label{sd1}
\eeqar
Here $ a^{\dagger}_{k^{\p}\uparrow} a_{k\downarrow}$ are the operators of $s$-electrons and $S^{-}(k^{\p}-k)$ is the spin lowering operator of $d$-or $f$ electrons ($S^{-}(q)=\sum_{k}a^*_{k+q\dwa}a_{k\upa}$) .

Another novelty of our calculation over the published calculations of electrical resistivity \cite{kasuya,mannari,weiss,mathon,mills,toru,ueda,kumari} is that it is manifestly beyond the Relaxation Time Approximation (RTA) which is taken into account in the memory function formalism \cite{singh,gotze} (our main tool in the current investigation) and full temperature evolution of the resistivity can be calculated whereas in the refs \cite{kasuya,mannari,weiss,mathon,mills,toru,ueda,kumari} resistivity is calculated either using the variational solution of the Block-Boltzmann equation or the iterative approximate method\cite{ziman,singh}. The problem with the Bloch-Boltzmann approach is that the full temperature evolution of resistivity is difficult to obtain analytically (only in low and hight temperature limits (say, with respect to the Debye temperature), the collision integral can be analytically simplified). Within the memory function formalism, we could analyse the full temperature evolution of resistivity rigorously and point out two regimes of interest: In the low temperature regime ($k_{B} T<<\mud$), we find an upturn in the resistivity and in the high temperature regime ($k_{B} T>>\mud$), we find that $\rho\propto T^\frac{3}{2}$. We compare our theory with the experimental data of ref. \cite{bachar} and find good agreement.
\section{Computational procedure using MF formalism}
In Kubo's linear response theory, the dynamical conductivity is given by 
\beqar
\sigma_{\mu\nu}(\om)= V\int_{0}^{\infty} dt e^{i\om t}\int_{0}^{\beta} d\lambda\la J_{\mu}(-i\hbar\lambda) J_{\nu}(t)\ra.  \label{eq1}
\eeqar
This is called the Kubo formula \cite{kubo,singh,gotze}. By using the Mori-Zwanzig projection operator technique the above Kubo formula can be rewritten in the following form\cite{singh,gotze}
\beqar
\sigma_{\mu\nu}(z)=i\frac{\om^2_{p}}{4\pi}\frac{1}{z+M_{\mu\nu}(z)}. \label{eq2}
\eeqar
Here $M_{\mu\nu}(z)$ is called the memory function and $z$ is the complex frequency ($z=\om+i\delta$). Thus the problem of computation of the dynamical conductivity boils down to the computation of the memory function $M_{\mu\nu}(z)$. Within the  G\"{o}tze-W\"{o}lfle approach the memory function is computed using the equation of motion method and a perturbative expansion of the memory function. All the technical details are given in refs.\cite{singh,gotze} here we outline the approach. It turns out that
\beqar
M(z)\simeq \frac{1}{z}(\frac{ne^2}{m})[\la\la \dot{J_{1}};\dot{J_{1}}\ra\ra_{z}-\la\la \dot{J_{1}};\dot{J_{1}}\ra\ra_{0}] \label{eq3}
\eeqar
where
\beqar
\dot{J_{1}}=-\frac{i}{\hbar}[J_{1},H] \label{eq4}
\eeqar
The total Hamiltonian is $H=H_{0}+H_{sd}$ and $H_{0}$ is the free electron unperturbed part and $H_{sd}$ is defined in eqn (\ref{sd1}). The double brackets are defined as 
\beqar
\la\la \hat{O}_{1} ;\hat{O}_{2} \ra\ra= i\frac{V}{\hbar} \int_{0}^{\infty} dt e^{iz t} \la [\hat{O}_{1}(t),\hat{O}_{2}(0)] \ra
\eeqar

Here $<...>$ means canonical ensemble average. The operator $\hat{O}(t)$ is in the Heisenberg representation $\hat{O}(t)=e^{iHt}\hat{O}(0)e^{-iHt}$. The current density operator is $J_{1}=\frac{1}{V}\sum_{k\sigma}ev_{k}a^{\dagger}_{k\sigma}a_{k\sigma}$ where $v_{k}=\frac{1}{\hbar}\frac{\pr \ep_{k}}{\pr k}$ and $V$ is the volume of the sample. With this information equation (\ref{eq4}) takes the form:
\beqar
\dot{J_{1}}=-i\frac{e}{\hbar V}\sum_{l,\sigma} \sum_{k^{\p}k}\bigg[v_{1}(l)a^{\dagger}_{l\sigma}a_{l\sigma},\frac{J}{N}\sum_{k,k^\p} \bigg\{ a^{\dagger}_{k^{\p}\uparrow} a_{k\downarrow}S^{-}(k^{\p}-k)+ a^{\dagger}_{k^{\p}\downarrow} a_{k\uparrow}S^{+}(k^{\p}-k)+\nonumber\\ (a^{\dagger}_{k^{\p}\uparrow}
a_{k\upa}-a^{\dagger}_{k^{\p}\downarrow} a_{k\dwa})S^{z}(k^{\p}-k)\bigg\}\bigg] \label{a1}
\eeqar
The current operator commutes with the unperturbed Hamiltonian, hence we are left with terms containing $H_{sd}$ which is treated as a perturbation.
Using Leibniz's bracket rule $[ab,c]=a\{b,c\}-\{a,c\}b$, the above expression reduces to
\beqar
\dot{J_{1}}=-\frac{i}{\hbar}\frac{eJ}{NV}\sum_{k^{\p}k}\bigg(v_{1}(k^\p)-v_{1}(k)\bigg)\bigg(a^{\dagger}_{k^{\p}\uparrow} a_{k\downarrow}S^{-}(k^{\p}-k)+a^{\dagger}_{k^{\p}\downarrow} a_{k\uparrow}S^{+}(k^{\p}-k)\bigg). \label{a2}
\eeqar

Define the correlator $\phi(z)=\la\la\dot{J_{1}};\dot{J_{1}}\ra\ra$: 
\beqar
\phi(z)&=&\frac{-e^2 J^2}{N^2\hbar^2 V^2} \sum_{k^{\p}k}\sum_{pp^\p}\bigg(v_{1}(k^\p)-v_{1}(k)\bigg)\bigg(v_{1}(p)-
v_{1}(p^\p)\bigg)\la\la a^{\dagger}_{k^{\p}\uparrow} a_{k\downarrow}S^{-}(k^{\p}-k) +\nonumber\\ &&~~~~~~~ a^{\dagger}_{k^{\p}\downarrow} a_{k\uparrow}S^{+}(k^{\p}-k) ~;~ a^{\dagger}_{p\uparrow} a_{p^\p\downarrow}S^{-}(p-p^\p) + a^{\dagger}_{p\downarrow} a_{p^\p\uparrow}S^{+}(p-p^\p) \ra\ra .
\label{c2}
\eeqar
Then the memory function (\ref{eq3}) can be written as $M(z)\simeq \frac{1}{z}(\frac{ne^2}{m})(\phi(z)-\phi(0))$. This is called the G\"{o}tze-W\"{o}lfle memory function approximation \cite{singh,gotze}. Now for the computation of memory function we need to compute the correlator($\phi(z)$)
\begin{equation}
\phi(z)=\la\la \dot{J_{1}};\dot{J_{1}}\ra\ra=i\frac{V}{\hbar}\int_{0}^{\infty} e^{izt}\la~[\dot{J_{1}}(t);\dot{J_{1}}(0)]~\ra dt . \label{c1}
\end{equation}
The correlation function $\phi(z)$ can be simplified to
 \beqar
 \phi(z)&=&\frac{-e^2 J^2}{N^2\hbar^2 V^2} \sum_{k^{\p}k}\sum_{pp^\p}\bigg(v_{1}(k^\p)-v_{1}(k)\bigg)\bigg(v_{1}(p)-
 v_{1}(p^\p)\bigg)\bigg\{\la\la a^{\dagger}_{k^{\p}\uparrow} a_{k\downarrow}S^{-}(k^{\p}-k);a^{\dagger}_{p\downarrow} a_{p^\p\uparrow}S^{+}(p-p^\p) \ra\ra\nonumber\\ &&~~~~~~~~~~~~~~~~+\la\la a^{\dagger}_{k^{\p}\downarrow} a_{k\uparrow}S^{+}(k^{\p}-k);a^{\dagger}_{p\uparrow} a_{p^\p\downarrow}S^{-}(p-p^\p) \ra\ra\bigg\},\label{c3}
 \eeqar
as the cross-terms of the form $\la\la a^{\dagger}_{k^{\p}\uparrow} a_{k\downarrow}S^{-}(k^{\p}-k);a^{\dagger}_{p\uparrow} a_{p^\p\downarrow}S^{-}(p-p^\p) \ra\ra$ vanish \cite{singh,gotze}. 
 We separate the function $\phi(z)$ into two sub functions $\phi_{1}(z)$ and $\phi_{2}(z)$ for simplification. The first function takes the form:

 \beqar
 \phi_{1}(z)&=&-i\frac{e^2J^2}{N^2\hbar^3 V} \sum_{k^{\p}k}\sum_{pp^\p}\bigg(v_{1}(k^\p)-v_{1}(k)\bigg)\bigg(v_{1}(p)-
 v_{1}(p^\p)\bigg)\int_{0}^{\infty} dt e^{izt} \la[ a^{\dagger}_{k^{\p}\uparrow}(t) a_{k\downarrow}(t)S^{-}(k^{\p}-k,t),\nonumber\\&&~~~~~~~~~~~~~~~~~~~~~~~~~~~~~~~~~~~a^{\dagger}_{p\downarrow} a_{p^\p\uparrow}S^{+}(p-p^\p)] \ra \label{c3a}
 \eeqar
 It is to be noted the impurity and conduction electron spin flip terms of the form of eqn (2) are incorporated in the commutator in the above equation (\ref{c3a}) that is $a^{\dagger}_{k^{\p}\uparrow}(t) a_{k\downarrow}(t)S^{-}(k^{\p}-k,t),a^{\dagger}_{p\downarrow} a_{p^\p\uparrow}S^{+}(p-p^\p)$ etc.
We write the time dependence of operators explicitly as $a^{\dagger}_{k^{\p}\uparrow}(t)=e^{\frac{i\ep_{k^\p}t}{\hbar}}a^{\dagger}_{k^{\p}\uparrow}(0)$ for $s$-band mobile electrons\footnote{As $a^{\dagger}_{k^{\p}\uparrow}(t)$ is in the Heisenberg representation, it should be written as  $a^{\dagger}_{k^{\p}\uparrow}(t)=e^{\frac{i\ep^{T}_{k^\p}t}{\hbar}}a^{\dagger}_{k^{\p}\uparrow}(0)$, where $\ep^{T}_{k^\p}$ is the eigenvalues of the total Hamiltonian $H=H_{0}+H_{sd}$. But we have replaced $\ep^{T}_{k^\p}$ with $\ep_{k^\p}$ which is the eigenvalue of the unperturbed or free electron Hamiltonian $H_{0}$. This approximation is valid as the perturbation $H_{sd}$ is assumed weaker (weak coupling limit of $J$) also refer to \cite{gotze}.}. For   $d-$ band density operators we write $S^{-}(k^{\p}-k,t)=e^{-i\om_{k^\p-k}t}S^{-}(k^{\p}-k,0)$. In the present case $\hbar\om_{k^\p-k}$ represents the spin flip energy of an excitation of the quasi localized of $d$ or $f$ electrons. Dispersion of the magnetic excitation created by operators $S^{-}(q)$ and $S^{+}(q)$ is assumed to be of the form $\hbar\om_{q}\propto q^2$  in the long wavelength limit which we use in the present calculation\cite{mannari}. Next on performing the time integration and applying anticommutating 
 Leibniz rule{\footnote{\{ab,cd\}=a\{b,c\}d-ac\{b,d\}+\{a,c \}db-c\{a,d\}b} to the Fermion operators in equation (\ref{c3a}) we obtain
 \beqar
 \phi_{1}(z)&=& C_{1} \sum_{k^{\p}k}\sum_{pp^\p}(\frac{1}{\frac{\ep^s_{k^\p}}{\hbar}-\frac{\ep^s_{k}}{\hbar}-\om_{k^\p-k}+z})\bigg(v_{1}(k^\p)-v_{1}(k)\bigg)\bigg(v_{1}(p)-
 v_{1}(p^\p)\bigg)\la( -a^{\dagger}_{k^{\p}\uparrow}a_{k\downarrow}a^{\dagger}_{p\downarrow} a_{p^\p\uparrow})\nonumber\\&& [S^{-}(k^{\p}-k),S^{+}(p-p^\p)]+\la\{ a^{\dagger}_{k^{\p}\uparrow}a_{k\downarrow},a^{\dagger}_{p\downarrow} a_{p^\p\uparrow}\}S^{+}(p-p^\p)S^{-}(k^{\p}-k) \ra.
 \label{c4}
 \eeqar 
Here $C_{1}=\frac{e^2J^2}{N^2\hbar^3 V}$. We write $\la a^{\dagger}_{k^{\p}\uparrow}  a_{k\downarrow} a^{\dagger}_{p\downarrow} a_{p^\p\uparrow} \ra= \la a^{\dagger}_{k^{\p}\upa} (\delta_{k p}-a^{\dagger}_{p\dwa}a_{k\dwa} ) a_{p^\p\upa} \ra$ and use bracket rule$^{ii}$ to solve factor $\la \{ a^{\dagger}_{k^{\p}\uparrow}a_{k\downarrow},a^{\dagger}_{p\downarrow} a_{p^\p\uparrow}\}\ra$. On simplifying, using the properties of delta functions $\delta_{k,p}$ and $\delta_{k^\p, p^\p}$, we get:
 \beqar
  \phi_{1}(z)&=&-~ C_{1} \sum_{k^{\p}k}(\frac{1}{\frac{\ep^s_{k^\p}}{\hbar}-\frac{\ep^s_{k}}{\hbar}-\om_{k^\p-k}+z})(v_{1}(k^\p)-v_{1}(k))^2\bigg[ f^s_{k^{\p}\upa}(1-f^s_{k\dwa})\la\{S^{-}(k^{\p}-k),S^{+}(k-k^\p)\} \ra\nonumber\\&&~~~~~~~~~~~~~~~+(f^s_{k^{\p}\upa}-f^s_{k\dwa})\la S^{+}(k-k^\p) S^{-}(k^{\p}-k) \ra \bigg] . \label{c5}
  \eeqar
  Here $ f^s_{k^\p\upa}=\la a^{\dagger}_{k^{\p}\uparrow}a_{k^{\p}\upa}\ra $ is the Fermi function of the $s$-band electrons. The spin density operators of $d$-band transforms the expression (\ref{c5}) to [refer to appendix A]
  
 \beqar
 \phi_{1}(z)&=& -~ C_{1} \sum_{k^{\p}k}(\frac{1}{\frac{\ep^s_{k^\p}}{\hbar}-\frac{\ep^s_{k}}{\hbar}-\om_{k^\p-k}+z})(v_{1}(k^\p)-v_{1}(k))^2\bigg[f^s_{k^\p\upa}(1-f^s_{k\dwa})\sum_{k_{d},k^{\p}_{d}} (f^{d}_{k_{d}\upa}-f^{d}_{k^{\p}_{d}\dwa})-\nonumber\\&&~~~~~~~~~~~~~~~~~~~~~~~~(f^s_{k\dwa}-f^s_{k^\p\upa}) \sum_{k_{d},k^{\p}_{d}}f^{d}_{k_{d}\upa}(1-f^d_{k^{\p}_{d}\dwa})\bigg] . \label{c6}
  \eeqar
   Similary write $ \phi_{2}(z)$ part from equation (\ref{c3}) :
   \beqar
   \phi_{2}(z)&=& -\frac{e^2J^2}{N^2\hbar^3 V^2} \sum_{k^{\p}k}\sum_{pp^\p}\bigg(v_{1}(k^\p)-v_{1}(k)\bigg)\bigg(v_{1}(p)-
   v_{1}(p^\p)\bigg)\la\la a^{\dagger}_{k^{\p}\downarrow} a_{k\uparrow}S^{+}(k^{\p}-k)~;\nonumber\\&&~~~~~~~~~~~~~~~~~~~a^{\dagger}_{p\uparrow} a_{p^\p\downarrow}S^{-}(p-p^\p) \ra\ra.  \label{c7}
   \eeqar
Again following the similar steps that are followed for the calculation of $\phi_{1}(z)$, we obtain expression for $\phi_{2}(z)$ as:
 \beqar
 \phi_{2}(z)&=&-~ C_{1} \sum_{k^{\p}k}(\frac{1}{\frac{\ep^s_{k^\p}}{\hbar}-\frac{\ep^s_{k}}{\hbar}-\om_{k^\p-k}-z})\bigg(v_{1}(k^\p)-v_{1}(k)\bigg)^2 \bigg[f^s_{k^\p\upa}(1-f^s_{k\dwa}) \sum_{k_{d},k^{\p}_{d}}(f^{d}_{k_{d}\upa}-f^{d}_{k^{\p}_{d}\dwa})-\nonumber\\&&~~~~~~~~~~~~~~~~~~~~~~~~(f^s_{k\dwa}-f^s_{k^\p\upa}) \sum_{k_{d},k^{\p}_{d}}f^{d}_{k_{d}\upa}(1-f^d_{k^{\p}_{d}\dwa})\bigg].  \label{c8}
  \eeqar
 We drop the spin notation in Fermi functions as there is no Zeeman splitting(no external and internal magnetic fields present). The total $\phi(z)$ takes the form:
 \beqar
 \phi(z)&=&-\frac{e^2J^2}{N^2\hbar^3 V}\sum_{k^{\p}k}(v_{1}(k^\p)-v_{1}(k))^2\bigg\{ f^s_{k^\p}(1-f^s_{k}) \sum_{k_{d},k^{\p}_{d}} (f^{d}_{k_{d}}-f^{d}_{k^{\p}_{d}})-(f^s_{k}-f^s_{k^\p})\times \nonumber\\&&~~~~~~~~~~~~~~~~~~~~~\sum_{k_{d},k^{\p}_{d}} f^{d}_{k_{d}}(1-f^d_{k^{\p}_{d}})\bigg\}  
 \bigg[ \frac{1}{\frac{\ep_{k^\p}}{\hbar}-\frac{\ep_{k}}{\hbar}-\om_{k^\p-k}+z} +\frac{1}{\frac{\ep_{k^\p}}{\hbar}-\frac{\ep_{k}}{\hbar}-\om_{k^\p-k}-z}  \bigg].\nonumber\\\label{c9}
 \eeqar
 \section{Computation of the Memory Function in the DC limit }
 Our aim is to determine the dynamical conductivity $\sigma(z)$ that depends on the Memory function, therefore writing $\phi(z)$ in terms of $M(z)$ using formula $M(z)=\frac{1}{z}\frac{m}{ne^2}(\phi(z)-\phi(0))$, we obtain 

\beqar
M(z)&=&-\frac{J^2 m}{N^2\hbar^3 n  V \om}\sum_{k^{\p}k}(v_{1}(k^\p)-v_{1}(k))^2\bigg\{ f^s_{k^\p}(1-f^s_{k})\sum_{k_{d},k^{\p}_{d}} (f^{d}_{k_{d}}-f^{d}_{k^{\p}_{d}})-(f^s_{k}-f^s_{k^\p}) \sum_{k_{d},k^{\p}_{d}}f^{d}_{k_{d}}(1-f^d_{k^{\p}_{d}})\bigg\}\nonumber\\&&   
\bigg[ \frac{1}{\frac{\ep_{k^\p}}{\hbar}-\frac{\ep_{k}}{\hbar}-\om_{k^\p-k}+z} +\frac{1}{\frac{\ep_{k^\p}}{\hbar}-\frac{\ep_{k}}{\hbar}-\om_{k^\p-k}-z}-\frac{1}{\frac{\ep_{k^\p}}{\hbar}-\frac{\ep_{k}}{\hbar}-\om_{k^\p-k}}-\frac{1}{\frac{\ep_{k^\p}}{\hbar}-\frac{\ep_{k}}{\hbar}-\om_{k^\p-k}}  \bigg]\nonumber\\ \label{m1}
\eeqar 
Where $M(z)=M(\om\pm i 0)=M^\p(\om)\pm i M^{\p\p}(\om)$. Here we are interested in the imaginary part of the memory function \cite{singh,gotze}. The use of identity $\lim_{\eta\rta 0} \frac{1}{a\mp i\eta}=\mathfrak{P}(\frac{1}{a})\pm i\pi\delta(a)$ transforms the expression (\ref{m1}){\footnote{$\lim_{\eta\rta 0}\frac{1}{\ep_{k^\p}-\ep_{k}-\om_{q}+\om \pm i\eta}=\mathfrak{P}(\frac{1}{\ep_{k^\p}-\ep_{k}-\om_{q}+\om})\mp i\pi \delta(\ep_{k^\p}-\ep_{k}-\om_{q}+\om)$}}} into delta function form. On comparing  imaginary part of the above expression, we get 
  
 \beqar
 M^{\p\p}(\om)&=&\frac{J^2 m\pi}{N^2\hbar^3 n  V \om} \sum_{k^{\p}k}(v_{1}(k^\p)-v_{1}(k))^2\{  f^s_{k^\p}(1-f^s_{k})\sum_{k_{d},k^{\p}_{d}} (f^{d}_{k_{d}}-f^{d}_{k^{\p}_{d}})-\nonumber\\&&(f^s_{k}-f^s_{k^\p}) \sum_{k_{d},k^{\p}_{d}}f^{d}_{k_{d}}(1-f^d_{k^{\p}_{d}})\} [\delta(\frac{\ep_{k^\p}}{\hbar}-\frac{\ep_{k}}{\hbar}-\om_{k^\p-k}+\om)-\delta(\frac{\ep_{k^\p}}{\hbar}-\frac{\ep_{k}}{\hbar}-\om_{k^\p-k}-\om)]. \label{m2} \nonumber\\&&
\eeqar 
Using the momentum conservation $\vec{k}^\p-\vec{k}=\vec{k}^{\p}_{d}-\vec{k}_{d}=\vec{q}$, write $\vec{k}^\p$ and $\vec{k}^{\p}_{d}$ in terms of $\vec{k}+\vec{q}$ and $\vec{k}_{d}+\vec{q}$. Also
write  $(v_{1}(k^\p)-v_{1}(k))^2=\frac{\hbar^2}{m^2}(\vec{k}^\p-\vec{k})^2$. To deal with the magnitude of $(\vec{k}^\p-\vec{k})$, i.e. $|\vec{k}^\p-\vec{k}|$ insert an integral $dq \delta (\vec{q}-|\vec{k}^\p-\vec{k}|)$ over $q$ into equation (\ref{m2}) which simplify the calculation greatly. Using the spatial isotropy in the present free electron case we can write $v^2=(v^2_{x}+v^2_{y}+v^2_{z})=3v^2_{x}$.  Converting sums into integrals for $k$ and $k^\p$ using $\frac{1}{V}\sum\rta\int\frac{ d^3k}{(2\pi)^3}$, the above equation can be written as
 \beqar
 M^{\p\p}(\om)&=&\frac{J^2\pi V }{3N^2 m n  }\int_{0}^{\infty}\frac{ dq}{\om} ~q^2\int_{0}^{\infty}  \frac{d^3k}{(2\pi)^3}\int_{0}^{\infty}\frac{d^3k^\p}{(2\pi)^3} \delta (\vec{q}-|\vec{k}^\p-\vec{k}|)F(f^s_{k},f^s_{k^\p},f^{d}_{k_{d}},f^d_{k^{\p}_{d}})\nonumber\\&&~~~~~~~~~~~~~~~~~~~~~[\delta(\ep_{k+q}-\ep_{k}-\hbar\om_{q}+\hbar\om)-\delta(\ep_{k+q}-\ep_{k}-\hbar\om_{q}-\hbar\om)]. \label{m3}
  \eeqar
Here, we write $F(f^s_{k},f^s_{k^\p},f^{d}_{k_{d}},f^d_{k^{\p}_{d}})$ as short hand notation for Fermi distribution function inside the  curly braces. Write $\int d^3k=4\pi \int k^2 dk$, $\int d^3k^\p=2\pi \int k^{\p 2}dk^\p\int_{0}^{\pi}\sin\theta d\theta$ (take  $k$ as pointing along the $z-$direction). Therefore $ M^{\p\p}(\om)$ takes the form

  \beqar
  M^{\p\p}(\om)&=&\frac{2J^2\pi V }{3N^2 m n  }\frac{(2\pi)^2}{(2\pi)^6}\int_{0}^{\infty} \frac{ dq}{\om} ~q^2\int_{0}^{\infty} k^2 dk \int_{0}^{\infty}k^{\p 2} dk^\p\int_{0}^{\pi}\sin\theta d\theta \delta (q-\sqrt{(k^{\p 2}+k^2-2k^{\p} k\cos\theta)})\nonumber\\&&
  \sum_{k_{d},k^{\p}_{d}}F(f^s_{k},f^s_{k^\p},f^{d}_{k_{d}},f^d_{k^{\p}_{d}})[\delta(\ep_{k+q}-\ep_{k}-\hbar\om_{q}+\hbar\om)-\delta(\ep_{k+q}-\ep_{k}-\hbar\om_{q}-\hbar\om)]. \label{m4}
  \eeqar 
  To simplify further, we shift momentum integral variables into energy variables $k^2=\frac{2 m\ep}{\hbar^2}$ and $dk=\frac{1}{\hbar}\sqrt{\frac{m}{2\ep}}d\ep$. On writing $\ep_{k}$ as $\ep$ and $\ep_{k^{\p}}$ as $\ep^\p$ changes the expression to
   \beqar
  M^{\p\p}(\om)&=&\frac{2J^2 V m^2 }{3N^2\hbar^6  n  }\frac{1}{(2\pi)^3}\int_{0}^{\infty} \frac{dq q^2}{\om}\int_{0}^{\infty} \sqrt{\ep} d\ep \int_{0}^{\infty} \sqrt{\ep^\p} d\ep^\p \int_{0}^{\pi}\sin\theta d\theta\times \nonumber\\&&~~~~~~~~~ \delta (q-\sqrt{2m}\sqrt{(\ep^\p+\ep-2\sqrt{\ep^\p \ep}\cos\theta)}) \sum_{k_{d}}F(f^s_{k},f^s_{k^\p},f^{d}_{k_{d}},f^d_{k^{\p}_{d}})\times \nonumber\\&&~~~~~~~~~~~~~~~~~~~~~~~~~[\delta(\ep_{k+q}-\ep_{k}-\hbar\om_{q}+\hbar\om)-\delta(\ep_{k+q}-\ep_{k}-\hbar\om_{q}-\hbar\om)]. \label{m5}
  \eeqar
 On performing the $\theta$ integral the above expression  (appendix B) reduces to the form 
  \beqar
  M^{\p\p}(\om)&=&\frac{1}{4\pi^3}\frac{J^2 V m^2 }{3N^2\hbar^6 n }\int_{0}^{q_{D}} \frac{dq q^2 q}{k^2_{s}\om }\int_{0}^{\infty} \sqrt{\ep} d\ep \int_{0}^{\infty} d\ep^\p\sqrt{\ep^\p}\{f^s_{k+q}(1-f^s_{k}) \sum_{k_{d}}(f^{d}_{k_{d}}-f^{d}_{k_{d}+q})-\nonumber\\&&~(f^s_{k}-f^s_{k+q})\sum_{k_{d}} f^{d}_{k_{d}}(1-f^d_{k_{d}+q})\}~ [\delta(\ep_{k+q}-\ep_{k}-\hbar\om_{q}+\hbar\om)-\delta(\ep_{k+q}-\ep_{k}-\hbar\om_{q}-\hbar\om)].\nonumber\\ \label{m6}
  \eeqar
  By using $f(x)\delta(x-a)=f(a)\delta(x-a)$ we remove $\ep_{k^\p}$ from the Fermi functions and  integrate over $\ep_{k^\p}$ which we simply write $\ep^\p$
  \beqar
  M^{\p\p}(\om)&=&p_{0}\int_{0}^{q_{D}}dq q^3 \int_{0}^{\infty} d\ep \frac{\sqrt{\ep}}{\om}\bigg[\sqrt{\ep_{k}+\hbar\om_{q}-\hbar\om}\sum_{k_{d}}\bigg\{f^s(\ep_{k}+\hbar\om_{q}-\hbar\om)(1-f^s(\ep_{k}))\times \nonumber\\&&\sum_{k_{d}}(f^{d}(\ep_{k_{d}})-f^{d}(\ep_{k_{d}+q}))-(f^s(\ep_{k})-f^s(\ep_{k}+\hbar\om_{q}-\hbar\om))
   \sum_{k_{d}}f^{d}(\ep_{k_{d}})(1-f^d(\ep_{k_{d}+q}))\bigg\}-\nonumber\\&&~~~~~~\sqrt{\ep_{k}+\hbar\om_{q}+\hbar\om}\bigg\{f^s(\ep_{k}+\hbar\om_{q}+\hbar\om)(1-f^s(\ep_{k}))\sum_{k_{d}}(f^{d}(\ep_{k_{d}})-f^{d}(\ep_{k_{d}+q}))-\nonumber\\&&~~~~~~~~~(f^s(\ep_{k})-f^s(\ep_{k}+\hbar\om_{q}+\hbar\om))\sum_{k_{d}} f^{d}(\ep_{k_{d}})(1-f^d(\ep_{k_{d}+q}))\bigg\}\bigg],
  \nonumber\\&& \label{m7}
 \eeqar
  where the prefactors $p_{0}=\frac{1}{4\pi^3}\frac{J^2 V m^2 }{3N^2\hbar^6  n q^2_{s} }$. Define $\ep^s_{k}=\ep$, $f^1_{d}=\sum_{k_{d}}\bigg(f^{d}(\ep_{k_{d}})-f^{d}(\ep_{k_{d}+q})\bigg)$ and $f^2_{d}= \sum_{k_{d}} f^{d}(\ep_{k_{d}})(1-f^d(\ep_{k_{d}+q}))$. With these definitions, we have
   \beqar
  M^{\p\p}(\om,T)&=&p_{0}\int_{0}^{q_{D}}dq q^3 \bigg \{\int_{0}^{\infty} d\ep \frac{\sqrt{\ep}}{\om} \bigg[\underbrace{\sqrt{\ep+\hbar\om_{q}-\hbar\om}f^s(\ep+\hbar\om_{q}-\hbar\om)(1-f^s(\ep))}_{term(T_{1})}-\nonumber\\&&~~~~~~~\underbrace{\sqrt{\ep+\hbar\om_{q}+\hbar\om}}_{term (T_{2})} \underbrace{f^s(\ep+\hbar\om_{q}+\hbar\om)(1-f^s(\ep))\bigg]}_{term(T_{2})}f^1_{d}(q)+\nonumber\\&&~~~~~~~~~~\int_{0}^{\infty} d\ep \frac{\sqrt{\ep}}{\om}\underbrace{\bigg[\sqrt{\ep+\hbar\om_{q}+\hbar\om}(f^s(\ep)-f^s(\ep+\hbar\om_{q}+\hbar\om))}_{term (T_{3})}\nonumber\\&&~~~~~~~~-\underbrace{\sqrt{\ep+\hbar\om_{q}-\hbar\om}(f^s(\ep)-f^s(\ep+\hbar\om_{q}-\hbar\om))\bigg]}_{term (T_{4})} f^2_{d}(q)\bigg\} . \label{m8}
  \eeqar
  This is important general expression of imaginary part of the Memory Function, which is valid for all frequencies and all temperature regimes. In what follows, we analyze the above expression in the D.C. limit and study the temperature dependence of the imaginary part of the memory function. For performing the limit $\om\rta0$, we rewrite the main result (equation \ref{m8}) in the following way:
  \beqar
   M^{\p\p}(\om)&=&p_{0}\int_{0}^{q_{D}}dq q^3 \bigg[ \int_{0}^{\infty}d\ep\sqrt{\ep}\times\nonumber\\&& \bigg(\underbrace{\frac{\sqrt{\ep+\hbar\om_{q}-\hbar\om}f^s(\ep+\hbar\om_{q}-\hbar\om)-\sqrt{\ep+\hbar\om_{q}+\hbar\om}f^s(\ep+\hbar\om_{q}+\hbar\om)}{\om}}_{Te1}\bigg)\nonumber\\&&\times(1-f^s(\ep))f^1_{d}(q)+\int_{0}^{\infty}d\ep\sqrt{\ep} \bigg(\underbrace{\frac{\sqrt{\ep+\hbar\om_{q}+\hbar\om}-\sqrt{\ep+\hbar\om_{q}-\hbar\om}}{\om}}_{Te2}\bigg)f^s(\ep)f^2_{d}(q)+\nonumber\\&&\int_{0}^{\infty}d\ep\sqrt{\ep} \bigg(\frac{\sqrt{\ep+\hbar\om_{q}-\hbar\om}f^s(\ep+\hbar\om_{q}-\hbar\om)-\sqrt{\ep+\hbar\om_{q}+\hbar\om}f^s(\ep+\hbar\om_{q}+\hbar\om)}{\om}\bigg)f^2_{d}(q)\bigg]. \nonumber\\&& \label{m8a}
  \eeqar
  
On performing the limit $\om\rta0$ for term (Te1) we have
\beqar
\frac{\pr Te1}{\pr \om}|_{\om=0}&=& -\hbar\frac{f^s(\ep+\hbar\om_{q}-\hbar\om)}{2\sqrt{\ep+\hbar\om_{q}-\hbar\om}}+\sqrt{\ep+\hbar\om_{q}-\hbar\om}\frac{\pr f^s(\ep+\hbar\om_{q}-\hbar\om)}{\pr \om}|_{\om=0}-\hbar\frac{f^s(\ep+\hbar\om_{q}+\hbar\om)}{2\sqrt{\ep+\hbar\om_{q}+\hbar\om}}\nonumber\\&&~~~~~~~~~~-\sqrt{\ep+\hbar\om_{q}+\hbar\om}\frac{\pr f^s(\ep+\hbar\om_{q}+\hbar\om)}{\pr \om}|_{\om=0}\nonumber\\
&=& -\hbar\frac{f^s(\ep+\hbar\om_{q})}{\sqrt{\ep+\hbar\om_{q}}}+2\hbar\beta\sqrt{\ep+\hbar\om_{q}} \frac{e^{\beta(\ep+\hbar\om_{q}-\mu_{s})}}{(e^{\beta(\ep+\hbar\om_{q}-\mu_{s})}+1)^2} \label{hte1}
\eeqar
and for term (Te2), we have
\beqar
\frac{\pr Te2}{\pr \om}|_{\om=0}&=&\frac{\hbar}{2\sqrt{\ep+\hbar\om_{q}+\hbar\om}}|_{\om=0}+\frac{\hbar}{2\sqrt{\ep+\hbar\om_{q}-\hbar\om}}|_{\om=0}=\frac{\hbar}{\sqrt{\ep+\hbar\om_{q}}}. \label{hte2}
\eeqar
Substituting the above expressions into eqn (\ref{m8a}) we obtain the memory function in the D.C. limit
 \beqar
M^{\p\p}(T)&=&p_{0}\hbar\int_{0}^{q_{D}}dq q^3 \bigg[ \int_{0}^{\infty}d\ep\sqrt{\ep}\bigg(-\frac{f^s(\ep+\hbar\om_{q})}{\sqrt{\ep+\hbar\om_{q}}}+2\beta \sqrt{\ep+\hbar\om_{q}} \frac{e^{\beta(\ep+\hbar\om_{q}-\mu_{s})}}{(e^{\beta(\ep+\hbar\om_{q}-\mu_{s})}+1)^2}\bigg)\times\nonumber\\&&~~~(1-f^s(\ep))f^1_{d}(q)+ \int_{0}^{\infty} \frac{d\ep\sqrt{\ep}}{\sqrt{\ep+\hbar\om_{q}}}f^s(\ep)f^2_{d}(q) + \int_{0}^{\infty}d\ep\sqrt{\ep}\bigg(-\frac{f^s(\ep+\hbar\om_{q})}{\sqrt{\ep+\hbar\om_{q}}}+\nonumber\\&&~~~~~~~2\beta \sqrt{\ep+\hbar\om_{q}} \frac{e^{\beta(\ep+\hbar\om_{q}-\mu_{s})}}{(e^{\beta(\ep+\hbar\om_{q}-\mu_{s})}+1)^2}\bigg)f^2_{d}(q)\bigg]\nonumber\\&&\label{memht}
\eeqar
There are a couple of reasonable assumptions which we would like to use to simplify the above expression: (1) The above expression can be simplified as $k_{B}T<<\mus$ (chemical potential for s-electrons) at temperature of interest ($\mus\simeq 10 eV$ and room temperature is $\sim \frac{1}{40} eV$). (2) $\hbar\om_{q}<<\mus$, that is, the energy scale of magnetic excitation (which is in $meV$) is much less than $\mus(\sim 10 eV)$. On implementing the second assumption in the Fermi function $f^s(\ep+\hbar\om_{q})=\frac{1}{e^{\beta(\ep+\hbar\om_{q}-\mu_{s})}+1}$ lead to $f^s(\ep)$ and the above expression becomes

\beqar
M^{\p\p}(T)&=&p_{0}\hbar\bigg[2\beta\int_{0}^{q_{D}}dq~ q^3 \int_{0}^{\infty}d\ep\sqrt{\ep} \sqrt{\ep+\hbar\om_{q}} f^s(\ep) (1-f^s(\ep))[(1-f^s(\ep))f^1_{d}(q)+f^2_{d}(q)]\nonumber\\&&~~~~-\int_{0}^{q_{d}}dq~ q^3 \int_{0}^{\infty}d\ep\sqrt{\ep}\frac{f^s(\ep)}{\sqrt{\ep+\hbar\om_{q}}}(1-f^s(\ep))f^1_{d}(q)\bigg] \label{mem12}
\eeqar
Next, on implementing the first assumption $k_{B}T<<\mu_{s}$, we notice that factors of the form $f^s(\ep) (1-f^s(\ep))$ are approximately like delta functions peaking at $\mus$. Thus the relevant range of the $\ep$ is around $\mus$ with width of order $k_{B}T$. Observing this fact we can write $\sqrt{\ep+\hbar\om_{q}}\simeq \sqrt{\ep}$ as $\hbar\om_{q}<<\mus$:
\beqar
M^{\p\p}(T)&=&p_{0}\hbar\bigg[2\beta\int_{0}^{\infty}d\ep\sqrt{\ep} \sqrt{\ep}~f^s(\ep) (1-f^s(\ep))\bigg((1-f^s(\ep))\underbrace{\int_{0}^{q_{d}}dq~ q^3f^1_{d}(q)}_{I_{1}(T)}+\nonumber\\&&\underbrace{\int_{0}^{q_{d}}dq~ q^3f^2_{d}(q)}_{I_{2}(T)}\bigg)-\int_{0}^{\infty}d\ep\sqrt{\ep}\frac{f^s(\ep)}{\sqrt{\ep}}(1-f^s(\ep))\underbrace{\int_{0}^{q_{d}}dq~ q^3f^1_{d}(q)}_{I_{1}(T)}\bigg] \label{new1}
\eeqar
~~~~~~~~~~~~~~~~~~~~~~~~~~~~~~~~~~~~~~~~~~~Or
\beqar
M^{\p\p}(T)&=&p_{0}\hbar\bigg[2\beta\int_{0}^{\infty}d\ep\ep~f^s(\ep) (1-f^s(\ep))\bigg((1-f^s(\ep))I_{1}(T)+I_{2}(T)\bigg)- \nonumber\\&&~~~~~~~~~~~\int_{0}^{\infty}d\ep\ f^s(\ep) (1-f^s(\ep))I_{1}(T) \label{new2}
\eeqar
Integrals over $\ep$ can be performed using the properties of delta functions $f^s(\ep) (1-f^s(\ep))\simeq \frac{1}{\beta}\delta(\ep-\mus)$:


\beqar
M^{\p\p}(T)&=&\frac{p_{0}\hbar}{\beta}\bigg[ (\beta\mu_{s}-1)I_{1}(T)+2\beta\mu_{s}I_{2}(T)\bigg] \label{new5}
\eeqar
As $\beta\mu_{s}>>1$, we get
\beqar
M^{\p\p}(T)&=&p_{0}\hbar\mu_{s}\bigg[ I_{1}(T)+2I_{2}(T)\bigg] \label{new6}
\eeqar
where  
\beqar
I_{1}(T)=\int_{0}^{q_{D}}dq~ q^3f^1_{d}(q), \label{new7}
\eeqar
and 
\beqar
I_{2}(T)=\int_{0}^{q_{D}}dq~ q^3f^2_{d}(q). \label{new7a}
\eeqar
The above simplified expression (eqn \ref{new6}) is our main result in the DC limit. Our next aim is to reduce the expression for $I_{1}(T)$ and $I_{2}(T)$. For this we take the long wavelength approximation (small $q$ expansion). It can be shown (refer to Appendix C) that $f^1_{d}(\ep_{d})$ in long wavelength limit $q\rta0$ can be written as
\beqar
f^1_{d}(\ep_{d})&=&\frac{Vq^2}{4\pi^2}\frac{\sqrt{2m}}{\hbar}\bigg[\beta \int_{0}^{\infty}\frac{d\ep_{d} \sqrt{\ep_{d}}e^{\beta(\ep_{d}-\mu_{d})}}{(e^{\beta(\ep_{d}-\mu_{d})}+1)^2}+\frac{2}{3}\beta^2 \int_{0}^{\infty} \frac{d\ep_{d} \ep_{d}^{\frac{3}{2}}e^{\beta(\ep_{d}-\mu_{d})}}{(e^{\beta(\ep_{d}-\mu_{d})}+1)^2}-\frac{4}{3}\beta^2\int_{0}^{\infty} \frac{d\ep \ep^{\frac{3}{2}}e^{2\beta(\ep_{d}-\mu_{d})}}{(e^{\beta(\ep_{d}-\mu_{d})}+1)^3}\bigg],\nonumber\\&& \label{memht5}
\eeqar
on substituting the above expression of $f^1_{d}(\ep_{d})$ into eqn  (\ref{new7}) we get
\beqar
I_{1}(T)&=& \frac{q^6_{D}}{6}\frac{V}{4\pi^2}\frac{\sqrt{2m}}{\hbar}\bigg[\beta \int_{0}^{\infty}\frac{d\ep_{d} \sqrt{\ep_{d}}e^{\beta(\ep_{d}-\mu_{d})}}{(e^{\beta(\ep_{d}-\mu_{d})}+1)^2}+\frac{2}{3}\beta^2 \int_{0}^{\infty} \frac{d\ep_{d} \ep_{d}^{\frac{3}{2}}e^{\beta(\ep_{d}-\mu_{d})}}{(e^{\beta(\ep_{d}-\mu_{d})}+1)^2}\nonumber\\&&~~~~~~~-\frac{4}{3}\beta^2\int_{0}^{\infty} \frac{d\ep \ep^{\frac{3}{2}}e^{2\beta(\ep_{d}-\mu_{d})}}{(e^{\beta(\ep_{d}-\mu_{d})}+1)^3}\bigg]. \label{new8a}
\eeqar
Similarly $f^2_{d}$ can be simplified (refer to appendix D) and the simplified expression of $f^2_{d}$ can be substituted into eqn (\ref{new7a}). The result is
\beqar
I_{2}(T)=\frac{V}{(2\pi)^2}\frac{(2m)^{\frac{3}{2}}}{\hbar^3}\int_{0}^{q_{D}}dq~ q^3\int_{0}^{\infty}\frac{d\ep_{d} \sqrt{\ep_{d}}e^{\beta(\ep_{d}-\mu_{d})}}{(e^{\beta(\ep_{d}-\mu_{d})}+1)^2} + I_{1}(T) ,     \label{nu3}
\eeqar
on substituting expressions of $I_{1}(T)$ and $I_{2}(T)$ into eqn (\ref{new6}) we have
\beqar
M^{\p\p}(T)&=&\frac{1}{12\pi^3}\frac{J^2 V^2 m^2 }{N^2\hbar^5  n q^2_{s} }\mu_{s}\bigg\{ \frac{3q^6_{D}}{6}\frac{1}{4\pi^2}\frac{\sqrt{2m}}{\hbar}\bigg(\beta \int_{0}^{\infty}\frac{d\ep_{d} \sqrt{\ep_{d}}e^{\beta(\ep_{d}-\mu_{d})}}{(e^{\beta(\ep_{d}-\mu_{d})}+1)^2}+\frac{2}{3}\beta^2\times \nonumber\\&&\int_{0}^{\infty} \frac{d\ep_{d} \ep_{d}^{\frac{3}{2}}e^{\beta(\ep_{d}-\mu_{d})}}{(e^{\beta(\ep_{d}-\mu_{d})}+1)^2}-\frac{4}{3}\beta^2\int_{0}^{\infty} \frac{d\ep \ep^{\frac{3}{2}}e^{2\beta(\ep_{d}-\mu_{d})}}{(e^{\beta(\ep_{d}-\mu_{d})}+1)^3}\bigg)+\nonumber\\&&~~~~~~~~~~\frac{q^4_{D}}{(2\pi)^2}\frac{(2m)^{\frac{3}{2}}}{2\hbar^3}\int_{0}^{\infty}\frac{d\ep_{d} \sqrt{\ep_{d}}e^{\beta(\ep_{d}-\mu_{d})}}{(e^{\beta(\ep_{d}-\mu_{d})}+1)^2}\bigg\}, \nonumber\\&& \label{nu4}
\eeqar
transforms the variables in all the integrands to  $x=\beta(\ep_{d}-\mu_{d})$:

\beqar
M^{\p\p}(T)&=&\frac{1}{12\pi^3}\frac{J^2 V^2 m^2 }{N^2\hbar^5  n q^2_{s} }\mu_{s}\bigg\{\frac{q^6_{D}}{8\pi^2}\frac{\sqrt{2m\mus}}{\hbar}\bigg(\frac{1}{\sqrt{\beta\mus}}\int_{-\beta \mud}^{\infty} dx\sqrt{x+\beta\mud}\frac{e^x}{(e^x+1)^2}+\nonumber\\&&
\frac{2}{3}\frac{1}{\sqrt{\beta\mus}}\int_{-\beta \mud}^{\infty} dx (x+\beta\mud)^\frac{3}{2}\frac{e^x}{(e^x+1)^2}-\frac{4}{3}\frac{1}{\sqrt{\beta\mus}}\int_{-\beta \mud}^{\infty} dx (x+\beta\mud)^\frac{3}{2}\frac{e^{2x}}{(e^x+1)^3}\bigg)\nonumber\\&&+\frac{q^4_{D}}{8\pi^2}\frac{(2m\mus)^{\frac{3}{2}}}{\hbar^3}\frac{1}{(\beta\mus)^{\frac{3}{2}}}\int_{-\beta \mud}^{\infty}dx\sqrt{x+\beta\mud}\frac{e^x}{(e^x+1)^2}\bigg\} \label{nu6}.
\eeqar

We write $\sqrt{2m\mus}=\hbar q_{s}$ and $(2m\mus)^\frac{3}{2}=\hbar^3 q^3_{s}$. The above expression attains the form
\beqar
M^{\p\p}(T)&=&\frac{1}{12\pi^3}\frac{J^2 V^2 m^2 }{N^2\hbar^5  n }\mu_{s}\bigg\{\frac{1}{8\pi^2}(\frac{q_{D}}{q_{s}})^6 q^5_{s}\bigg(\frac{1}{\sqrt{\beta\mus}}\int_{-\beta \mud}^{\infty} dx \sqrt{x+\beta\mud}\frac{e^x}{(e^x+1)^2}+\nonumber\\&&
\frac{2}{3}\frac{1}{\sqrt{\beta\mus}}\int_{-\beta \mud}^{\infty} dx (x+\beta\mud)^\frac{3}{2}\frac{e^x}{(e^x+1)^2}-\frac{4}{3}\frac{1}{\sqrt{\beta\mus}}\int_{-\beta \mud}^{\infty} dx (x+\beta\mud)^\frac{3}{2}\frac{e^{2x}}{(e^x+1)^3}\bigg)\nonumber\\&&+\frac{1}{8\pi^2}(\frac{q_{D}}{q_{s}})^4 \frac{q^5_{s}}{(\beta\mus)^{\frac{3}{2}}}\int_{-\beta \mud}^{\infty}dx\sqrt{x+\beta\mud}\frac{e^x}{(e^x+1)^2}\bigg\}. \label{nu7}
\eeqar

This is our final simplified expression(after implementing the above mentioned assumptions 1 and 2). Temperature dependence of the imaginary part of memory function gives the temperature dependence of resistivity $\rho(T)=\frac{m}{ne^2}\frac{1}{\tau(T)}=\frac{m}{ne^2} M^{\p\p}(T)$\cite{singh,gotze}. The expression (\ref{nu7}) is plotted for various values of $\mud$ in figure (1a) and for various values of $q_{D}$ in figure (1b). We notice low temperature upturn (in (a) and (b)) and high temperature $T^\frac{3}{2}$ behaviour in figure (1c)(refer section 4.2 for details).

\begin{figure}
	\centering
\includegraphics[width=4.555cm]{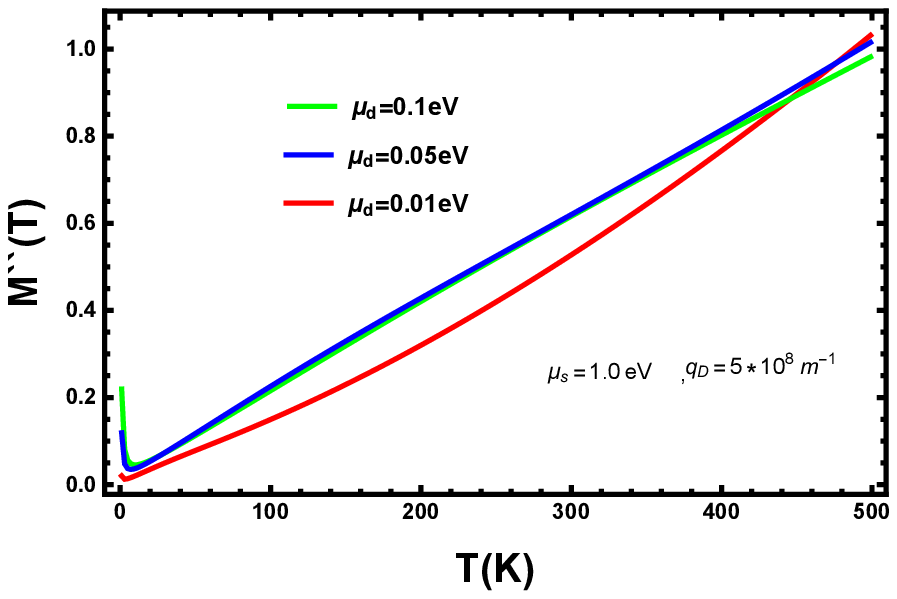}
\includegraphics[width=4.55cm]{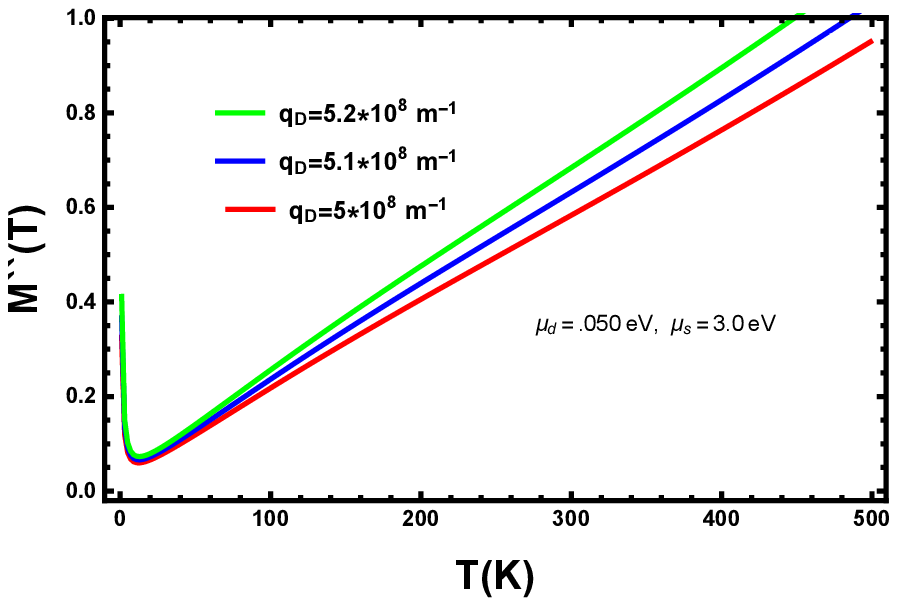}
\includegraphics[width=4.55cm]{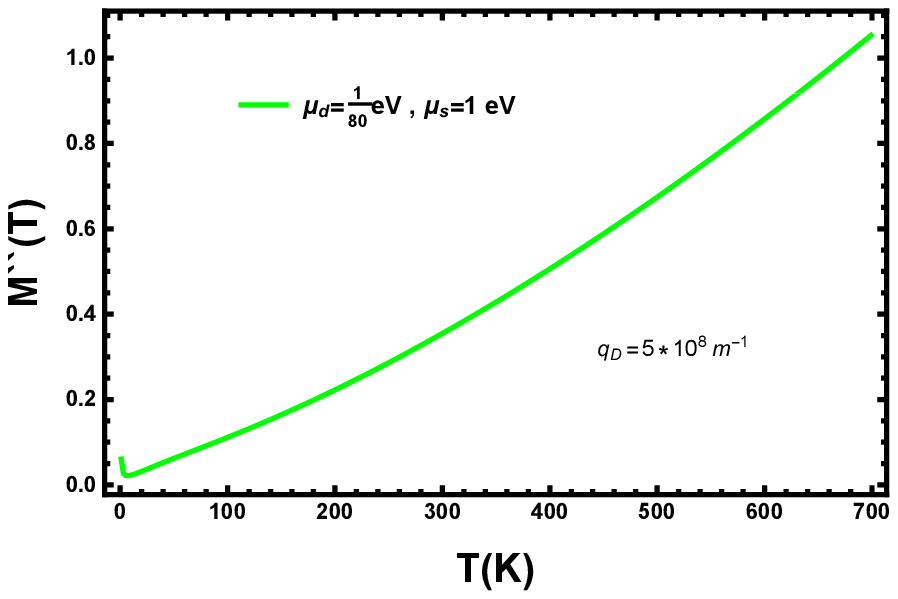}
\caption{(a) $M^{\p\p}(T)$ for various values of $\mud$. (b) $M^{\p\p}(T)$ for various values of $q_{D}$. (c) High temp. behaviour of $M^{\p\p}(T)$. $M^{\p\p}(T)\propto T^\frac{3}{2}$ in high temperature limit $k_{B}T>>\mud$.}
\label{fig:foobar}
\end{figure}


\section{Analysis of the general expression in special cases:}
\subsection{Low temperature limit ($k_{B}T<<\mu_{d}$)}
In this temperature limit we have $\beta\mud>>1$ thus the general expression (\ref{nu7}) transforms to 
\beqar
M^{\p\p}(T)&\simeq&\frac{1}{12\pi^3}\frac{J^2 V^2 m^2 }{N^2\hbar^5  n }\mu_{s}\bigg\{\frac{1}{8\pi^2}(\frac{q_{D}}{q_{s}})^6 q^5_{s}\bigg(\frac{\sqrt{\beta\mud}}{\sqrt{\beta\mus}}\int_{-\beta\mud}^{\infty} dx \frac{e^x}{(e^x+1)^2}+\frac{2}{3}\frac{(\beta\mus)^\frac{2}{3}}{\sqrt{\beta\mus}}\int_{-\beta\mud}^{\infty} dx \frac{e^x}{(e^x+1)^2} \nonumber\\&&~~~~~-\frac{4}{3}\frac{(\beta\mus)^\frac{2}{3}}{\sqrt{\beta\mus}} \int_{-\beta\mud}^{\infty} dx \frac{e^{2x}}{(e^x+1)^3}\bigg)+
\frac{1}{8\pi^2}(\frac{q_{D}}{q_{s}})^4 \frac{q^5_{s}}{(\beta\mus)^{\frac{3}{2}}}\sqrt{\beta\mud}\int_{-\beta\mud}^{\infty} dx \frac{e^x}{(e^x+1)^2}~, \nonumber\\&&
\eeqar
where we replaced $\sqrt{x+\beta\mud}\simeq \sqrt{\beta\mud}$ as $\beta\mud>>1$ and $x\sim 1$ due to exponentially damped function of the form $\frac{e^x}{(e^x+1)^2}$  in the integrands. With further rearrangements the above expression further simplifies to
\beqar
M^{\p\p}(T)&\simeq&\frac{1}{12\pi^3}\frac{J^2 V^2 m^2 }{N^2\hbar^5  n }\mu_{s}\bigg\{\frac{1}{8\pi^2}(\frac{q_{D}}{q_{s}})^6 q^5_{s}\bigg(\sqrt{\frac{\mud}{\mus}}\int_{-\beta\mud}^{\infty} dx \frac{e^x}{(e^x+1)^2}+\nonumber\\&&\frac{2}{3}\frac{(\mud)^{\frac{3}{2}}}{\sqrt{\mus}}\frac{1}{k_{B}T} \int_{-\beta\mud}^{\infty} d x \bigg\{\frac{e^x}{(e^x+1)^2}-2\frac{e^{2x}}{(e^x+1)^3}\bigg\}+\nonumber\\&&~~~~~~~~~~~~~\frac{1}{8\pi^2}(\frac{q_{D}}{q_{s}})^4 \frac{q^5_{s}\sqrt{\mud}}{(\mus)^{\frac{3}{2}}} k_{B}T\int_{-\beta\mud}^{\infty} dx \frac{e^x}{(e^x+1)^2}.
\eeqar
In the low temperature limit, the dominating term is the middle one with prefactor proportional to $\frac{1}{T}$. Neglecting the subdominating terms the memory function in low temperature limit reduces to
\beqar
M^{\p\p}(T\rta0) \sim  \frac{1}{T} f_{s}(T), ~~~~~~~  f_{s}(T)=\int_{-\beta\mud}^{\infty} d x \bigg\{\frac{e^x}{(e^x+1)^2}-2\frac{e^{2x}}{(e^x+1)^3}\bigg\},
\eeqar
where $f_{s}(T)$ is a slowly varying function\footnote{We have checked the relative variation of $f_{s}(T)$ as compared to $\frac{1}{T}$ and found that relative variation of $f_{s}(T)$ is very small.} of temperature. So, in the low temperature limit resistivity displays an upturn, as seen in figure (1a).  An important point to be noted here is that the divergence in our case is of the form of power law instead of the logarithmic divergence in the original Kondo problem. The reason behind this difference is that we treated d or f electrons as quasi-localized (away from half-filling) instead of fully localized ones\cite{patrik}. This is one of our important result.

\subsection{High temperature limit ($k_{B}T>>\mu_{d}$)}
In high temperature limit we have $\beta\mud<<1$. In this limit  expression from (\ref{nu7}) changes to
\beqar
M^{\p\p}(T)&\simeq&\frac{1}{12\pi^3}\frac{J^2 V^2 m^2 }{N^2\hbar^5  n }\mu_{s}\bigg\{\frac{1}{8\pi^2}(\frac{q_{D}}{q_{s}})^6 q^5_{s}\bigg( \frac{1}{\sqrt{\beta\mus}}\int_{0}^{\infty} dx \sqrt{x} \frac{e^x}{(e^x+1)^2}+\nonumber\\&&~~~~~~~~~~~~~~~~~~~\frac{2}{3}\frac{1}{\sqrt{\beta\mus}}\int_{0}^{\infty} dx (x)^\frac{2}{3} \bigg[\frac{e^x}{(e^x+1)^2}-2 \frac{e^{2x}}{(e^x+1)^3}\bigg]\bigg)+\nonumber\\&&~~~~~~~~~~~~~~~~~~~~\frac{1}{8\pi^2}(\frac{q_{D}}{q_{s}})^4 \frac{q^5_{s}}{(\beta\mus)^{\frac{3}{2}}}\int_{0}^{\infty} dx \sqrt{x} \frac{e^x}{(e^x+1)^2} \bigg\}.
\eeqar
By direct computation we notice that the last term in the above expression is many order of magnitude larger than the first two terms. Thus, 

\beqar
M^{\p\p}(k_{B}T>>\mu_{d})&\sim & C~ T^\frac{3}{2} \int_{0}^{\infty} dx \sqrt{x} \frac{e^x}{(e^x+1)^2}\sim 0.536~ C~ T^\frac{3}{2} \nonumber\\&& 
M^{\p\p}(k_{B}T>>\mu_{d})\sim T^\frac{3}{2}.
\eeqar
where prefactor  $C= \frac{1}{96\pi^5}\frac{J^2 V^2 m^2 }{N^2\hbar^5  n }(\frac{q_{D}}{q_{s}})^4\frac{q^5_{s}}{\sqrt{\mus}}$. Thus, in high temperature limit the memory function scales as $M^{\p\p}(k_{B}T>>\mu_{d})\sim T^\frac{3}{2}$. This is also observed in figure (1c).

\section{Comparison  with experimental data}

In this section we compare our theory with the experimental data. For comparison we consider Kondo-like behaviour observed in nano-scale granular aluminum samples \cite{bachar}.
\begin{figure}[h!]
	\centering
	\includegraphics[width=10cm]{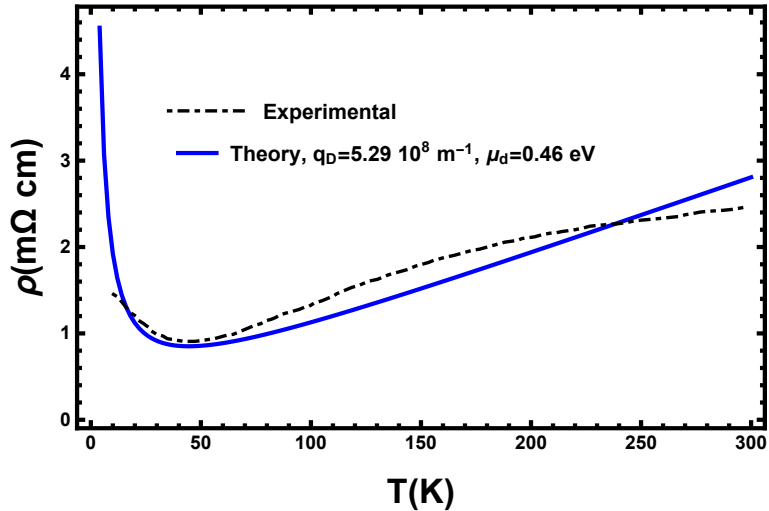}
	\caption{(a) Resistivity as a function of temperature compared with experimental data.... }
	\label{fig:foobar}
\end{figure}
 Resistivity of nano-scale granular aluminum samples was measured in reference\cite{bachar}. Kondo-like behaviour was observed in  the temperature dependence of resistivity. Resistivity shows low temperature upturn and a minimum around $T_{m}\simeq 40K$, and then it shows negative curvature at higher temperature ($T>T_{m}$). The experimental data in figure 2 of ref.\cite{bachar} is reproduced here in figure 2 (dotted line). In the experimental paper it is argued that such a resistivity behaviour originates from spin-flip scattering of conduction electrons by local magnetic moments which are possibly located at the metal oxide interface. The physical explanation given in the experimental paper is reasonable as our theory is in good agreement with the data (figure 2). In the present theory we have spin-flip scattering of conduction electrons off the quasi-localized $d$ or $f$ electrons. The DC resistivity is computed using the present theory $\rho(T)=\frac{m}{ne^2}\frac{1}{\tau(T)}=\frac{m}{ne^2} M^{\p\p}(T)$ takes the form
\beqar
\rho(T)&=&(\frac{m}{ne^2})\frac{1}{96\pi^5}\frac{J^2 V^2 m^2 }{N^2\hbar^5  n }\mu_{s}q^5_{s}\bigg\{(\frac{q_{D}}{q_{s}})^6 \bigg(\frac{1}{\sqrt{\beta\mus}}\int_{-\beta \mud}^{\infty} dx \sqrt{x+\beta\mud}\frac{e^x}{(e^x+1)^2}+\nonumber\\&&
\frac{2}{3}\frac{1}{\sqrt{\beta\mus}}\int_{-\beta \mud}^{\infty} dx (x+\beta\mud)^\frac{3}{2}\frac{e^x}{(e^x+1)^2}-\frac{4}{3}\frac{1}{\sqrt{\beta\mus}}\int_{-\beta \mud}^{\infty} dx (x+\beta\mud)^\frac{3}{2}\frac{e^{2x}}{(e^x+1)^3}\bigg)\nonumber\\&&+\frac{1}{8\pi^2}(\frac{q_{D}}{q_{s}})^4 \frac{1}{(\beta\mus)^{\frac{3}{2}}}\int_{-\beta \mud}^{\infty}dx\sqrt{x+\beta\mud}\frac{e^x}{(e^x+1)^2}\bigg\}, \label{rho1}
\eeqar
and it does show an up-turn at lower temperature, and as the temperature is raised it passes through a minima $(T_{m}\simeq 38 K)$ and then increase monotonically (figure 2). For the comparison of theory and experiment we take $\mus=11.5 eV$, and lattice constant $a=4.05 \AA$ (both for metal Aluminum). We take $q_{D}$ and $\mud$ as our fitting parameters. The best fit value is obtained for $q_{D}=5.29\times10^{8} m^{-1}$ and $\mud= 0.46 eV$. From the figure 2, it is clear that the theory developed here is in reasonable agreement with the experimental data. By comparing the magnitudes of $\mud\simeq 0.46~eV$ and $\mus\simeq 11.5~eV$ we notice that the s-electrons form a bigger Fermi surface, and d-electrons form a smaller Fermi surface, as expected from our theoretical considerations.

\section{Conclusion}
The calculation of DC resistivity through the calculation of the memory function formalism ($\rho(T)=\frac{m}{ne^2}\frac{1}{\tau(T)}=\frac{m}{ne^2} M^{\p\p}(T)$) for the Kondo lattice Hamiltonian (or $s$-$d$ Hamiltonian) is presented. We used the W\"{o}lfle-G\"{o}tze approximation to compute the memory function. The scattering of conduction electrons via the quasi-localized $f$ or $d$ electrons is taken into account by treating the $H_{s-d}$ part of Hamiltonian as a perturbation. Dispersion of spin excitations is taken to be of the form $\hbar\om_q = c_m q^2$. We find that the D.C. resistivity shows low temperature ($k_{B}T<<\mu_{d}$) power law up-turn and high temperature ($k_{B}T>>\mu_{d}$) $T^\frac{3}{2}$ scaling.

 \appendix 
 \section*{Appendices:}
  \section{ Average of spin density operators of localized electrons}
 The commutator of spin density operators is written as:
  \beqar 
  \sum_{k^{\p}k}\la[S^{-}(k^{\p}-k),S^{+}(k-k^\p)] \ra &=& \la[S^{-}(q),S^{+}(-q)] \ra\nonumber\\ 
  &=&\sum_{k} \la [a^{\dagger}_{k^\p\dwa} a_{k\upa},a^{\dagger}_{k\upa}a_{k^\p\dwa}]\ra \nonumber\\&& \label{ap1}
    \eeqar
    Here we set $k^{\p}-k=q$, and to treat $d$ electrons as quasi-localized we write $S^{+}$ and $S^{-}$ in terms of Fermi functions ($S^{-}(q)=\sum_{k}a^*_{k+q\dwa}a_{k\upa}$). The anticommutation property simplifies the eqn (\ref{ap1}) to
   \beqar
  \sum_{k^{\p} k}\la[S^{-}(k^{\p}-k),S^{+}(k-k^\p)] \ra &=&  \sum_{k,q} \la a^{\dagger}_{k^\p\dwa}\{a_{k\upa},a^{\dagger}_{k\upa}\}a_{k^\p\dwa} \ra-\la a^{\dagger}_{k\upa}  \{ a^{\dagger}_{k^\p\dwa},a_{k^\p\dwa}\}a_{k\upa} \ra\nonumber\\
  &=&\sum_{k,q}( f^{d}_{k+q\dwa}-f^{d}_{k\upa}) \label{ap2}
  \eeqar
  We use $f^d_{k^{\p}\dwa}=\la a^{\dagger}_{k^\p\dwa} a_{k^\p\dwa}\ra$ notation to differentiate Fermi function of $d$-band electrons from that of $s$-band electrons. The other factor in eqn (\ref{c5}) is:

  \beqar
  \sum_{k^{\p} k}\la  S^{+}(k-k^\p) S^{-}(k^{\p}-k)\ra&=& \la S^{+}(-q)S^{-}(q)\ra=\sum_{k+q}\la a^{\dagger}_{k\upa}a_{k+q\dwa} a^{\dagger}_{k+q\dwa} a_{k\upa}\ra\nonumber\\
  &=& \sum_{k,q} f^{d}_{k\upa}(1-f^d_{k+q\dwa}).\nonumber\\&& \label{ap3}
  \eeqar

    \section{$\theta$ integral solution}
In the presence of Fermi factors of the form $f^s_{k^\p}(1-f^s_{k})$ and at ordinary temperature  $k_{B}T<<\mus$($\sim$eV), one can replace $\ep$ and $\ep^{\p}$ inside the square root by $\mus$  for $s$ electrons ($\mus=\frac{\hbar^2 q^2_{s}}{2m}$) where $q_{s}$ is Fermi wavevector for $s$-electrons:
   \beqar
   \int_{0}^{\pi}\sin\theta d\theta \delta (q-\sqrt{2m}\sqrt{(\ep^\p+\ep-2\sqrt{\ep^\p \ep}\cos\theta)})&\simeq&\int_{0}^{\pi}\sin\theta d\theta \delta (q-2\sqrt{m\ep(1-\cos\theta)})\nonumber\\
   &\simeq&\int_{0}^{\pi}\sin\theta d\theta \delta (q-\sqrt{2}q_{s}\sqrt{(1-\cos\theta)})\nonumber\\ \label{th1}
   \eeqar
   Put $x=1-\cos\theta$ and define $\xi =q_{s}\sqrt{2x}$ and the limit of the integral changes to 0 and $2q_{s}$ (note that $0<q<q_{s})$. The integral becomes
    \beqar
   \int_{0}^{\pi}\sin\theta d\theta \delta (q-\sqrt{2m}\sqrt{(\ep^\p+\ep-2\sqrt{\ep^\p \ep}\cos\theta)})&\simeq& \int_{0}^{2k_{s}} \frac{\xi d\xi}{q^2_{s}}\delta(q-\xi)\simeq\frac{q}{q^2_{s}}.  \label{th2}
   \eeqar

   \section{Expansion of $f^1_{d}(q)$  }
   \beqar
   f^1_{d}(q)&=&\sum_{k_{d}}[ f^{d}(\ep_{k_{d}})-f^d(\ep_{k^{\p}_{d}})] \label{fd0}
   \eeqar 
   The Taylor's expansion for small ($ q \rta 0 $) gives
   \beqar
   f^1_{d}(q)&=& \sum_{k_{d}}[ f^{d}(\ep_{k_{d}})- f^{d}(\ep_{k_{d}})-q\frac{\pr f^{d}(\ep_{k^{\p}_{d}})}{\pr q}|_{q=0}-\frac{q^2}{2!}\frac{\pr^2 f^{d}(\ep_{k^{\p}_{d}})}{\pr q^2}|_{q=0}-\frac{q^3}{3!}\frac{\pr^3 f^{d}(\ep_{k^{\p}_{d}})}{\pr q^3}|_{q=0}...].\nonumber\\  \label{fd1}
   \eeqar
   on converting summation into integrals, we get
   \beqar
   f^1_{d}(q)&=&-\frac{V}{(2\pi)^2}\int_{0}^{\infty}k^2_{d} dk_{d}\int_{0}^{\pi}\sin\theta d\theta\bigg[q\frac{\pr f^{d}(\ep_{k^{\p}_{d}})}{\pr q}|_{q=0}+\frac{q^2}{2!}\frac{\pr^2 f^{d}(\ep_{k^{\p}_{d}})}{\pr q^2}|_{q=0}+\frac{q^3}{3!}\frac{\pr^3 f^{d}(\ep_{k^{\p}_{d}})}{\pr q^3}|_{q=0}...\bigg],\nonumber\\  \label{fd2}
   \eeqar
   We have Fermi function $f^{d}(\ep_{k^{\p}_{d}},\theta)=\frac{1}{e^{\beta[\frac{\hbar^2 q^2}{2m}+\frac{\hbar^2 k_{d}^2}{2m}+\frac{\hbar^2 k_{d} q \cos\theta}{m}-\mu_{d}]}+1}$. For simplification, we put $\alpha=\beta(\frac{\hbar^2 k_{d}^2}{2m}-\mu_{d})$, $\eta=\beta\frac{\hbar^2}{2m}$ and $\gamma=\beta\frac{\hbar^2 k_{d} }{m}$. The Fermi function set to
   \beqar
   f^d(q,\alpha,\eta,\gamma,\theta)=\frac{1}{e^{[\alpha+\eta q^2+\gamma q \cos\theta]}}~,~~~~\frac{\pr f^d(\alpha,\gamma,\theta)}{\pr q}|_{q=0}=-\frac{e^{\alpha}\gamma \cos\theta }{(e^{\alpha}+1)^2},\nonumber\\  \label{fd3}
   \eeqar
   similarly
   \beqar
   \frac{\pr^2  f^d(\alpha,\eta,\gamma,\theta)}{\pr q^2}|_{q=0}=-\frac{e^{\ep_{d}-\mu_{d}}}{(e^{\ep_{d}-\mu_{d}}+1)^2}[2 \eta+\gamma^2 \cos^2\theta]+\frac{2\gamma^2e^{\beta(\ep_{d}-\mu_{d})}  \cos^2\theta}{(e^{\ep_{d}-\mu_{d}}+1)^3} , \label{fd4}
   \eeqar
   the third derivative becomes
   \beqar
   \frac{\pr^3  f^d(\alpha,\eta,\gamma,\theta)}{\pr q^3}|_{q=0}=\frac{12\eta e^{2\alpha}\gamma \cos\theta}{(e^{\alpha}+1)^3}-\frac{6\eta e^{\alpha}\gamma \cos\theta}{(e^{\alpha}+1)^2}-\frac{6 e^{3\alpha}\gamma^3 \cos^3\theta}{(e^{\alpha}+1)^4}+\frac{6 e^{2\alpha}\gamma^3 \cos^3\theta}{(e^{\alpha}+1)^3}-\frac{ e^{\alpha}\gamma^3 \cos^3\theta}{(e^{\alpha}+1)^2},\nonumber\\ \label{fd5}
   \eeqar   
  We substitute derivative terms of $f^{d}(\ep_{k^{\p}_{d}})$ from eqn (\ref{fd3}),(\ref{fd4}) and (\ref{fd5}) in the expression (\ref{fd2}) and perform $\theta$ integration. Thus replacing $\alpha$, $\eta$ and $\gamma$ with their respective terms we obtain
    \beqar
   f^1_{d}(\ep_{d})&=&V\frac{q^2}{4\pi^2}\frac{\sqrt{2m}}{\hbar}\bigg[\beta \int_{0}^{\infty}\frac{d\ep_{d} \sqrt{\ep_{d}}e^{\beta(\ep_{d}-\mu_{d})}}{(e^{\beta(\ep_{d}-\mu_{d})}+1)^2}+\frac{2}{3}\beta^2 \int_{0}^{\infty} \frac{d\ep_{d} \ep_{d}^{\frac{3}{2}}e^{\beta(\ep_{d}-\mu_{d})}}{(e^{\beta(\ep_{d}-\mu_{d})}+1)^2}-\frac{4}{3}\beta^2\int_{0}^{\infty} \frac{d\ep \ep^{\frac{3}{2}}e^{2\beta(\ep_{d}-\mu_{d})}}{(e^{\beta(\ep_{d}-\mu_{d})}+1)^3}\bigg].\nonumber\\&&~~\label{fd7}
  \eeqar
     \section{Term $f^2_{d}(q)$ expansion}
    The Fermi function of $d$-band electrons $f^2_{d}(q)$ is
    \beqar
    f^2_{d}(q)&=&\sum_{k_{d}} f^{d}(\ep_{k_{d}})(1-f^d(\ep_{k^{\p}_{d}}) \label{fd2a}
    \eeqar 
    The Taylor's expansion for small $q$ expands the Fermi function in the form
    \beqar
    f^2_{d}(q)&=&\sum_{k_{d}} f^{d}(\ep_{k_{d}})\bigg(1- f^{d}(\ep_{k_{d}})-q\frac{\pr f^{d}(\ep_{k^{\p}_{d}})}{\pr q}|_{q=0}-\frac{q^2}{2!}\frac{\pr^2 f^{d}(\ep_{k^{\p}_{d}})}{\pr q^2}|_{q=0}-....\bigg)\nonumber\\
    &=& \sum_{k_{d}}\bigg[ f^{d}(\ep_{k_{d}})\bigg(1- f^{d}(\ep_{k_{d}})\bigg)-qf^{d}(\ep_{k_{d}})\frac{\pr f^{d}(\ep_{k^{\p}_{d}})}{\pr q}|_{q=0}-\frac{q^2}{2!}f^{d}(\ep_{k_{d}})\frac{\pr^2 f^{d}(\ep_{k^{\p}_{d}})}{\pr q^2}|_{q=0}-...\bigg].\nonumber\\&& \label{fd2b}
    \eeqar
   On converting sum into integration
    \beqar
    f^2_{d}(\ep_{k_{d}})&=& V\bigg[\frac{1}{(2\pi)^2} \int_{0}^{\infty}k^2_{d} dk_{d}f^{d}(\ep_{k_{d}})(1-f^d(\ep_{k_{d}}))\int_{0}^{\pi}\sin\theta d\theta-\frac{q^2}{2!(2\pi)^2} \int_{0}^{\infty}k^2_{d} dk_{d}f^{d}(\ep_{k_{d}})\times  \nonumber\\&&~~~~\int_{0}^{\pi}\sin\theta d\theta\frac{\pr^2 f^{d}(\ep_{k^{\p}_{d}})}{\pr q^2}|_{q=0}-...... \bigg],\label{fd2c}
    \eeqar
    which can further be written in terms of energy
    \beqar
   f^2_{d}(\ep_{k_{d}})&=& \frac{V}{(2\pi)^2}\frac{(2m)^{\frac{3}{2}}}{\hbar^3}\int_{0}^{\infty}\frac{d\ep_{d} \sqrt{\ep_{d}}e^{\beta(\ep_{d}-\mu_{d})}}{(e^{\beta(\ep_{d}-\mu_{d})}+1)^2}+\frac{Vq^2}{4\pi^2}\frac{\sqrt{2m}}{\hbar}\bigg[\beta \int_{0}^{\infty}\frac{d\ep_{d} \sqrt{\ep_{d}}e^{\beta(\ep_{d}-\mu_{d})}}{(e^{\beta(\ep_{d}-\mu_{d})}+1)^2}+\nonumber\\&&~~~~~~~~~~\frac{2}{3}\beta^2 \int_{0}^{\infty} \frac{d\ep_{d} \ep_{d}^{\frac{3}{2}}e^{\beta(\ep_{d}-\mu_{d})}}{(e^{\beta(\ep_{d}-\mu_{d})}+1)^2}-\frac{4}{3}\beta^2\int_{0}^{\infty} \frac{d\ep \ep^{\frac{3}{2}}e^{2\beta(\ep_{d}-\mu_{d})}}{(e^{\beta(\ep_{d}-\mu_{d})}+1)^3}\bigg]. \label{fd2d}
    \eeqar


\end{document}